\documentclass{article}
\usepackage[english]{babel}
\usepackage[a4paper,top=2cm,bottom=2cm,left=3cm,right=3cm,marginparwidth=1.75cm]{geometry}
\usepackage{times}
\usepackage{soul}
\usepackage{color}
\usepackage{graphicx}
\usepackage{ucs}
\usepackage[utf8x]{inputenc}
\usepackage{amsmath}
\usepackage{amstext}
\usepackage{amssymb}
\usepackage{slashed}
\usepackage{mathtools}
\usepackage{enumerate}
\usepackage[normalem]{ulem}
\usepackage{makeidx}
\usepackage[breaklinks=true,colorlinks=false,hidelinks]{hyperref}
\usepackage[english]{cleveref}
\usepackage[font=footnotesize]{caption}
\usepackage{subcaption}
\usepackage{array}
\usepackage[dvipsnames]{xcolor}
\usepackage{dirtytalk}
\usepackage{setspace}
\usepackage{cite}
\usepackage{amsmath}
\usepackage{graphicx}
\spacing{1.5}

\title{\textbf{Donor-acceptor discrete optical emission in 2D perovskites}}

\author
{Setatira Gorji$^{1,\dagger}$, Marie Krecmarova$^{1,\dagger}$, Alejandro Molina$^{1}$, Maria C. Asensio$^{2,3}$\\ Andrés F. Gualdrón-Reyes$^{4,5}$, Jesús Rodríguez-Romero$^{4,6}$, Hamid Pashaei-Adl$^{1}$\\ Rodolfo Canet-Albiach$^{1}$, Luca Schio$^{7}$, Massimo Tormen$^{7}$, Luca Floreano$^{7}$,\\ Iván Mora-Seró$^{4}$, Juan Martínez Pastor$^{1,3}$, Juan Francisco Sánchez-Royo$^{1,3}$\\ and Guillermo Muñoz Matutano$^{1,*}$
\\
\normalsize{$^{1}$Instituto de Ciencia de Materiales,
Universidad de Valencia (ICMUV), 46071 Valencia, Spain}\\
\normalsize{$^{2}$Instituto de Ciencia de Materiales de Madrid (ICMM), CSIC 28049 Madrid, Spain}\\
\normalsize{$^{3}$MATINÉE: CSIC Associated Unit (ICMM-ICMUV), Universidad de Valencia, Valencia, Spain}\\
\normalsize{$^{4}$Institute of Advanced Materials (INAM), Universitat Jaume I, Avenida de Vicent Sos Baynat,}\\
\normalsize{s/n, 12071, Castelló de la Plana, Spain}\\
\normalsize{$^{5}$Facultad de Ciencias, Instituto de Ciencias Químicas, Isla Teja, Universidad Austral de Chile,}\\
\normalsize{5090000, Valdivia, Chile.}\\
\normalsize{$^{6}$Facultad de Química, Universidad Nacional Autónoma de México, Circuito Exterior s/n, C.U.,}\\
\normalsize{Coyoacán, 04510 Mexico City, Mexico.}\\
\normalsize{$^{7}$ CNR-IOM, Laboratorio TASC, 34149 Trieste, Italy.}\\
\\
\normalsize{$^\ast$Corresponding author E-mail:  guillermo.munoz@uv.es}\\
\normalsize{$\dagger$ These authors contributed equally}
}

\begin{document}
\maketitle

\begin{abstract}
Two-dimensional (2D) van der Waals nanomaterials have attracted considerable attention for potential use in photonic and optoelectronic applications in the nanoscale, due to their outstanding electrical and optical properties, differing from their bulk state. Currently, 2D perovskite belonging to this group of nanomaterials is widely studied for a wide range of optoelectronic applications. Thanks to their excitonic properties, 2D perovskites are also promising materials for photonics and nonlinear devices working at room temperature. Nevertheless, strong excitonic effects can reduce the photocurrent characteristics when using thinner perovskites phases. In this work, we present solid experimental evidence for the presence of single donor-acceptor pair optical transitions in 2D Lead Halide Perovskites, characterized by sub meV linewidths ($\simeq 120 \mu eV$) and long decay times (5-8 ns). Micro-photoluminescence evidence is supported by detailed Photoemission measurements, and a model simulation. The association of Phenethylammonium with Methylammonium cations, the latter molecule being only present in 2D Halide Perovskites with thicker phases $n \geq 2$, has been identified as the source of the donor-acceptor pair formation, corresponding to the displacement of lead atoms and their replacement by methylamonium. Our seminal study of discrete Donor-Acceptor Pair (DAP) sharp and bright optical transitions in 2D Lead Halide Perovskites opens new routes to implement DAP as the carrier sources for novel designs of optoelectronic devices with 2D perovskites, and will foster the development of future outstanding properties in non-linear quantum technologies.

\end{abstract}

\section*{Introduction}

Research on perovskites has experienced rapid increase and visibility during the last decade due to their use in high performance and low cost solar cells, which represent a potential alternative to the commercially available Silicon photovoltaic wafers \cite{vinattieri2021halide}. Since the discovery of the high light-energy conversion efficiency published in 2009 \cite{kojima2009organometal}, the use of perovskite materials has been demonstrated in a large number of photonics devices owing to their interesting properties related to charge transport, excitons and photoexcited carriers generation, bandgap tunability and exciton binding energy in halide perovskites (HPs), and ferroelectric properties and magnetic field effects in oxide perovskites\cite{schmidt2021roadmap}. This rapid growth in academic activity has merged with a second intense field of research---the study of 2D monolayer semiconductors. The isolation of mechanically exfoliated graphene in 2004 has boosted the engineering of an interesting new set of high-quality 2D semiconductor samples that can be prepared at a low cost and without specialised equipment. This is the case for 2D Lead Halide Perovskites (LHPs) \cite{blancon2020semiconductor, zhu2020low}, which have recently been analysed more closely because of their potential integration in high-efficiency photovoltaics and photonic devices \cite{cao20152d, dou2017emerging, ricciardulli2021emerging, krahne2021two}.

As a 2D van der Waals family material, layered perovskites show enhanced stability and structural tunability \cite{leng2020bulk, blancon2020semiconductor, ricciardulli2021emerging}, but with a particular soft lattice and a dynamically disordered structure. Its crystal lattice typically consists of single or multiple organic–inorganic hybrid phases arranged into a Ruddlesden–Popper structure \cite{stoumpos2016ruddlesden}. In contrast to 3D perovskites, the semiconducting inorganic layer is passivated with insulating organic cations, forming a 2D natural quantum-well structure where the electronic excitations are confined in the inorganic layers \cite{leng2020bulk}. Its optical emission properties can be controlled by the n value, which labels the number of inorganic monolayers building the quantum well structure (from n = 1, 2, 3 to ≈∞) \cite{ricciardulli2021emerging, leng2020bulk}. Single layer 2D perovskites show strong Coulomb interaction, due to the effect of the quantum confinement and the weak dielectric screening. Consequently, excitonic states for 2D HPs with a small n number are characterized with binding energies up to several hundreds of meV \cite{blancon2020semiconductor}, which returns excitonic stability even at room temperature \cite{lekina2019excitonic} and following the emergence of large fine structure energy shifts \cite{do2020bright, Canet-Albiach2022Revealing}. 

The intense excitonic features that are present in these small n value 2D HPs play an important role in the development of new and promising optoelectronic devices. At first, the high excitonic binding energy characteristics of small n values 2D HPs reduces the overall photocurrent generation for solar cell applications \cite{green2014emergence}. Thus, for these purposes, larger n values samples (n = 3) with smaller excitonic binding energies are conventionally used \cite{zhao2018donor}. However, there is a different carrier strategy to produce high photocurrent generation in 2D HPs with small n values. The dissociation of excitons into Donor-Acceptor pairs (DAP) could be used to drive new device designs with small n value 2D HPs \cite{zhao2018donor}. The charge transfer process between excitons and DAP in n = 1 samples with small organic cations has been reported. However, the effect has not been observed when using Phenethylammonium (PEA) organic cations as layer spacers, or when using larger n values. Interestingly, a second study reveals that Donor-Acceptor states can be used to tune the excitonic binding energy in 2D hybrid layered perovskite samples \cite{passarelli2020tunable}, which is an important strategy to gain control of the performance of the final device. In a different direction, there is an open debate about the identification of the optical transitions that are present in the emission spectra of 2D LHPs, where extrinsic effects (e.g., DAP) are suggested as possible states producing red-shifted optical emission  \cite{kahmann2020extrinsic}. Therefore, the study of DAP states in 2D LHPs is revealed as an important area of research, where many important fundamental questions should be answered in detail, in order to enable the design and engineering of future 2D optoelectronic devices.

In this work, we measured and identified a single DAP carrier optical recombination that provides very narrow and long-lived optical transitions in 2D LHPs based materials at low temperature. By means of spatial and time resolved micro-photoluminiscence ($\mu$-PL), we recorded optical transitions with narrow linewidths, as low as $\simeq 120 \mu eV$ (full width half maximum, FWHM), and time decays as long as 8.6 ns. We identify the origin of these transitions as the DAP recombination that is created by exchange of methylammonium (MA) molecule and Pb atoms present in 2D HPs samples with phase thickness $n = 2$, leading to optical recombination in a broad spectral range between two excitonic transitions (i.e., free and trap exciton). This important claim is supported with strong experimental evidence by means of $\mu$-Raman spectra and high-resolution photoemission spectroscopy measurements. Our experimental effort is accompanied with predictions from a simple model to calculate the density of states (DOS) as a function of DAP binding energy, which is built up by the attractive Coulombic interaction between their net charges. Our work represents an important step forward in our understand of the nature of the DAP state in 2D LHPs, its spectral features, and its relationship with the selected cation spacer and the layer thickness n value. This has important consequences for the design of new optoelectronic architectures. Moreover, the ability to isolate single DAP states can potentially be used to study advanced polaritonic strategies with very sharp and bright optical transitions in a high non-linear material, with binding energy/DAP radius tunability, and hence to stimulate the use of 2D LHP semiconductors as promising materials to produce new alternatives to be considered in the actual quantum material database\cite{su2021perovskite, munoz2019emergence, delteil2019towards}.

\section*{Results}

\begin{figure}[h!]
	\centering
	\includegraphics[scale=0.45]{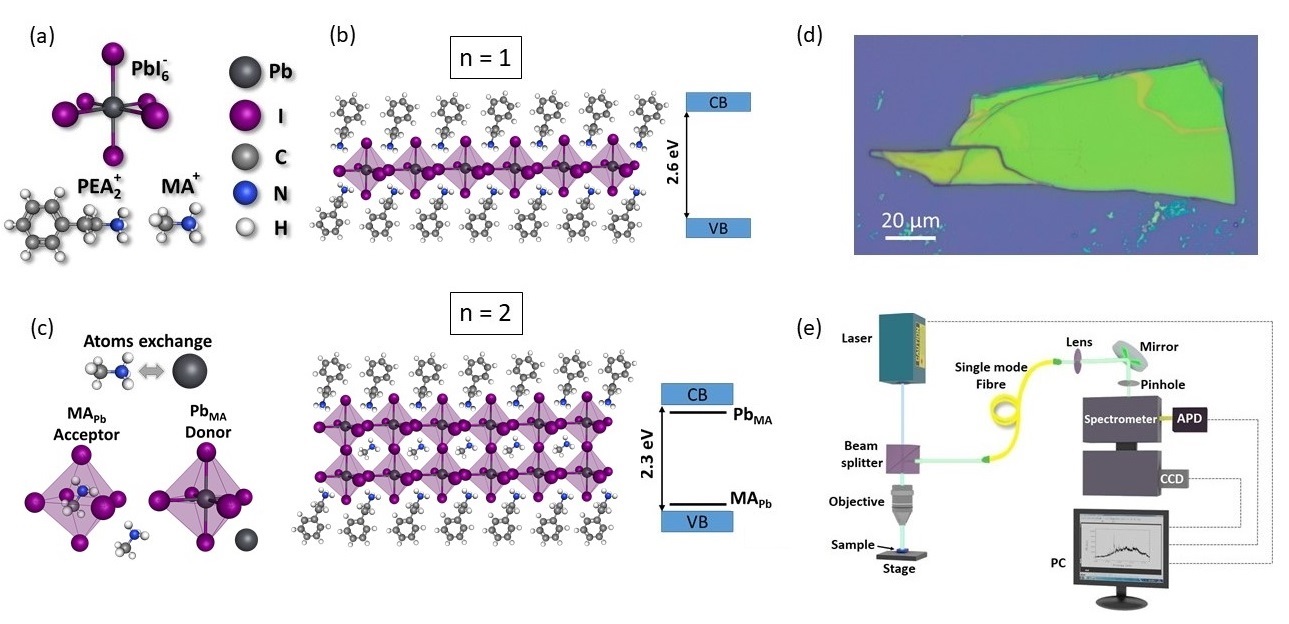}
	\caption{(a) 2D perovskites are composed of inorganic octahedral (PbI$_6$)$^-$, short organic methylammonium (MA)$^+$ cation, and long organic Phenethylammonium (PEA$_2$)$^+$ cation.(b) Crystal structure of a monolayer of 2D perovskite PEA$_2$PbI$_4$ crystal with a quantum well thickness of n=1, where a single octahedral sheet is sandwiched by a PEA$_2$ organic spacer. The inset image shows its bandgap with values taken from \cite{chen20182d}. (c) Crystal structure of a monolayer of 2D perovskite PEA$_2$MAPb$_2$I$_7$ crystal with a quantum well thickness of n=2, consisting of two octahedral sheets intercalated with short MA molecules, sandwiched by a PEA$_2$ organic spacer. The inset image shows its bandgap with values taken from \cite{chen20182d} and a donor-acceptor pair of Pb$_{MA}$ and MA$_{Pb}$ created by the MA and Pb atom exchanges \cite{xiao2016defect}. (d) Molecularly thin exfoliated flake of 2D perovskites with n=1 with tens of µm lateral size. (e) Scheme of the confocal micro-photoluminescence ($\mu$-PL) spectroscopy system with a steady state and time-resolved detection. A blue excitation laser beam with wavelength of 450 nm is focused through a beam splitter and an objective with NA=0.5 to the focal plane of 2D perovskite flake is used to achieve micro-localization. Steady state $\mu$-PL is detected by a charge-coupled device (CCD) camera and time-resolved $\mu$-PL by the avalanche photodiode (APD). 
}
	\label{Fig_1}
\end{figure}

\subsection*{Synthesis and preparation}
     Two different phases of 2D perovskites single crystals with the lowest quantum well thickness of n=1 and n=2 were synthesized  (see Materials and Methods for more details). The crystal lattice consists of an inorganic octahedral layer that is sandwiched by long organic cations, forming 2D quantum well structures (see Figure \ref{Fig_1}.a-c). The general chemical formula for 2D perovskites is R$_2$A$_{n-1}$B$_n$X$_{3n+1}$,  where R$_2$ is a long organic cation (phenethylammonium, butylammonium, ethylammonium, ...), A is a tiny organic cation (methylammonium (MA), formamidinium (FA), ...), B is a metal inorganic cation (e.g., lead, tin, and so on), and X is a halide inorganic anion (i.e., chloride, bromide, or iodide)  \cite{ricciardulli2021emerging, stoumpos2016ruddlesden,xiao2021layer,seitz2020exciton,zhao2018donor}. Figure \ref{Fig_1}(a-c) shows the crystal structures of both PEA$_2$PbI$_4$ (n=1) and PEA$_2$MAPb$_2$I$_7$ (n=2) phases. They are made up of a single layer (n=1) and a double layer (n=2) of perovskite inorganic octahedral flakes composed by (PbI6)$^-$ anions sandwiched by long organic (PEA2)$^+$ cations, which serve as spacers. The inorganic double layer is intercalated by short organic (MA)$^+$ cations for n = 2. Only the n = 1 phase has a well-defined structure, whereas small inclusions of another n phase are formed for a higher inorganic flake thickness $n > 1$ \cite{li2018fabrication}. 
     Synthesized crystals can be easily exfoliated down to a single monolayer, thanks to the relatively weak van der Waals interlayer coupling.
     Exfoliation reduces the occurrence of unwanted hybrid phase formation \cite{li2018fabrication} and leads to a strictly flat crystal orientation to the substrate \cite{dhanabalan2019simple}, which may suppress the substrate's lattice imperfections. In this study, synthesized single crystals of both phases were exfoliated onto SiO$_2$/Si substrates using scotch tape (see Materials and Methods for more details). Exfoliated single-crystal flake (n = 1) with a typical lateral size of tens to hundreds of micrometers and a thickness of units to tens of nanometers is shown in Figure \ref{Fig_1}.d.
     We used a confocal microscope to measure $\mu$-photoluminescence ($\mu$-PL) and time-resolved $\mu$-PL ($\mu$-TRPL), as shown in Figure \ref{Fig_1}.e. In our set-up, the samples are held in the cold finger of a closed cycle He cryostat with the possibility to tune temperature between 4-300 K (see Materials and Methods for more details) to investigate the optical characteristics of a large number of exfoliated flakes of 2D perovskites single crystals PEA$_2$PbI$_4$ (n=1) and PEA$_2$MAPb$_2$I$_7$ (n=2) At 4 K. (100 $\mu$eV at 532 nm)
     
\begin{figure}[h!]
	\centering
    \includegraphics[scale=0.3]{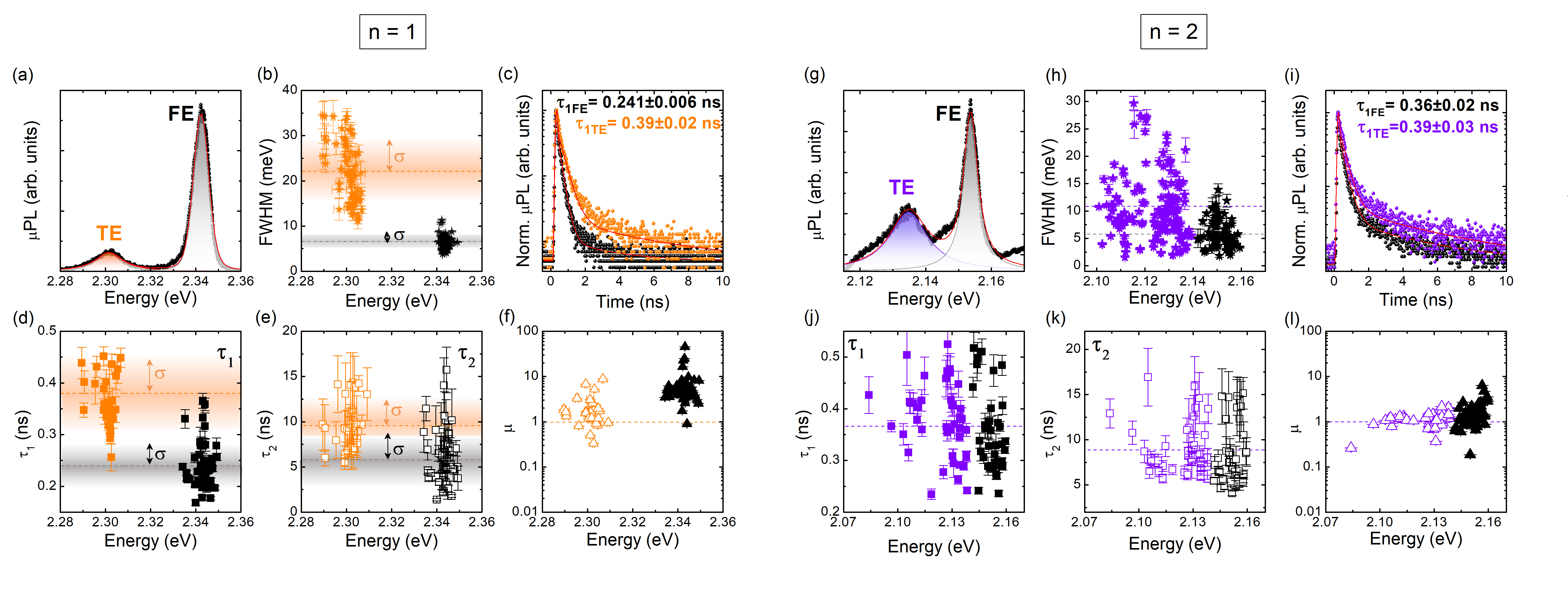}
	\caption{General photoluminescence (PL) and time resolved photoluminescence (TRPL) characteristics at low temperature (4 K). An example of measured PL spectra for phase (a) n=1 and (g) n=2. FE corresponds to free exciton, and TE corresponds to self-trapped exciton. Statistics of FE and TE peak linewidth full width at half maximum (FWHM) as a function of emission energy of FE and TE for phase (b) n=1 and (h) n=2. Example of TRPL of TE and FE on a 10 ns time scale for phase (c) n=1 and (i) n=2. Statistics of FE and TE, fast decay ($\tau_1$) for phases (d) n=1 and (j) n=2, and long decay ($\tau_2$) for phases (e) n=1 and (k) n=2. Statistics of the relative intensity between both fast and slow radiative channels of FE and TE, ($\mu$) as a function of emission energy for phases (f) n=1 and (l) n=2. }
	\label{Fig_2}
\end{figure}
     
     \subsection*{Micro-photoluminescence and time decays}
     
     Figure \ref{Fig_2}.a displays a typical optical spectrum for the n=1 phase samples. It is characterized by two main contributions of the emission PL spectra. The FE peak corresponds to the well-known free exciton recombination \cite{kahmann2020extrinsic, li2019surface}. However, the exciton may recombine by another process, such as intrinsic optical transitions (e.g.,self-trapped excitons (TE), biexcitons, triexcitons, or phonon replicas) or extrinsic optical transitions as excitons bound to defects \cite{blancon2020semiconductor}. As will be documented later on, TE transition has a higher FWHM than FE for both sample phases (n = 1, n = 2). The self-trapped exciton could be the most likely origin of the TE peak because electron-phonon interactions broaden the emission peak \cite{lekina2019excitonic, li2019surface}. We have fit the measured $\mu$-PL spectra of many different n = 1 and n = 2 flakes. Figure \ref{Fig_2}.b shows the evolution of the FWHM as a function of the emission energy for both FE and TE. FE (Black star) shows a mean FWHM = 6.5 meV with standard deviation $\sigma$ = 1.48 meV. TE (Orange star) data returns a mean FWHM = 22.1 meV and $\sigma$ = 6.8 meV. When analyzing FE and TE recombination in samples with n = 1, both transitions are clearly differentiated by their spectral characteristics. Figure \ref{Fig_2}.c shows a $\mu$-TRPL measurement example for both FE and TE transitions present in the n = 1 phase. TRPL is characterized by two decay channels (characterized with fast $\tau_{1}$ and slow $\tau_{2}$ decay times). FE shows a first fast decay time in the order of 0.241$\pm$0.006 ns and TE of 0.393$\pm$0.019 ns. Figures \ref{Fig_2}.d-e show a statistical sampling of the decay time measurement for both fast and slow components. The mean-value of FE $\tau_{1}$ is 0.24 ns, with a $\sigma$ = 0.04 ns; while for TE is 0.38 ns, with a $\sigma$ = 0.07 ns. Additionally, the residual long $\tau_{2}$ shows mean-value of 6 ns for FE transition, with $\sigma$ = 2.8 ns; while for TE it reaches a mean value of 9.5 ns, with $\sigma$ = 2.7 ns (Figure \ref{Fig_2}.e). Finally, we analyzed the relative intensity between both fast and slow radiative channels. To do this, we define new parameter ($\mu$) to compare the weight of the emitted light from both radiative channels. Hence, we define it as $\mu=\frac{I_{F}(T)}{I_{S}(T)}$, where $I_{F/S}(T)=\int_{t=0}^{t=T} a_{1}exp^{\frac{-t}{\tau_{1}/\tau_{2}}}\cdot dt$, and $T$ is the laser period. In this sense, the $\mu$ value will describe situations where the fast decay dominates ($\mu > 1$), both channels have similar contribution ($\mu \sim 1$), or where the slow decay dominates ($\mu < 1$). As is shown in the plot in figure \ref{Fig_2}.f, $\mu > 1$ for almost all cases, both for FE (filled black triangles) and TE (open orange triangles). In conclusion, in samples with phase n= 1, the fast decay time (~$\simeq$ hundreds of ps) always dominates over the long decay time ($\simeq$ ns).
     
     We now present a similar analysis carried out in thicker samples with phase n = 2. Figure \ref{Fig_2}.g shows an example of the $\mu$-PL spectra of a representative sample. First, and as before, we identify two main transitions. Following similar arguments as described earlier, these are labeled as FE and TE optical transitions. The increase of the thickness of the inorganic component red shifts their optical emission. The FE transition is shifted almost 200 meV towards low energy concerning its n = 1 counterpart. Figure \ref{Fig_2}.h shows FE (Black star) and TE (Violet star) FWHM values as a function of the emission peak energy, where the FE/TE mean FWHM is 5.7/10.4 meV, with $\sigma$ = 2.5/5.8 meV. Here, we clearly identify larger linewidth dispersion than in the previous statistical analysis, with n = 1. We tentatively associate this effect to the more complex electronic scenario present in the n = 2 phase. Next, we analyzed time decays. Figure \ref{Fig_2}.i shows one example of the FE and TE $\mu$-TRPL measurements, displaying the same double component of for both FE and TE peaks as in the n = 1 phase. Figures \ref{Fig_2}.j-k show the statistics of the fast ($\tau_{1}$) and slow ($\tau_{2}$) decay times for both FE and TE transitions. The mean $\tau_{1}$ value of FE is 0.35 ns, with a $\sigma$ = 0.07 ns; while for TE is 0.38 ns, with a $\sigma$ = 0.07 ns. The mean $\tau_{2}$ value of FE is 8.4 ns, with a $\sigma$ = 3.5 ns; while for TE is 8.8 ns, with a $\sigma$ = 2.8 ns. The principal difference between n = 1 and n = 2 $\mu$-TRPL is the higher observed dispersion for the n = 2 phase, although here mean values are slightly larger than previously. This larger dispersion reinforces the claim that a more complex scenario is present in this thicker phase. Finally, and as before, we analyzed the relative intensity of both radiative channels with the parameter $\mu$. Figure \ref{Fig_2}.i shows the dispersion of $\mu$ for the phase n = 2. As is shown, almost all values are situated close to $\mu \sim 1$ or $\mu > 1$. This means that the fast channel still dominates the optical emission in both  FE (filled black triangles) and TE (open violet triangles). In a small fraction of cases, the contribution of the slow component dominates ($\mu < 1$). This more complex behaviour adds to the previous findings related to the larger dispersion found for both FWHM, and $\tau_{1}$ and $\tau_{2}$ decays. Our study on the transient characteristics of the photoluminescence clearly identifies higher complexity when analyzing the n = 2 phase than the previous and simpler dynamical situation described in the n = 1 phase. 

\begin{figure}[h!]
	\centering
	\includegraphics[scale=0.33]{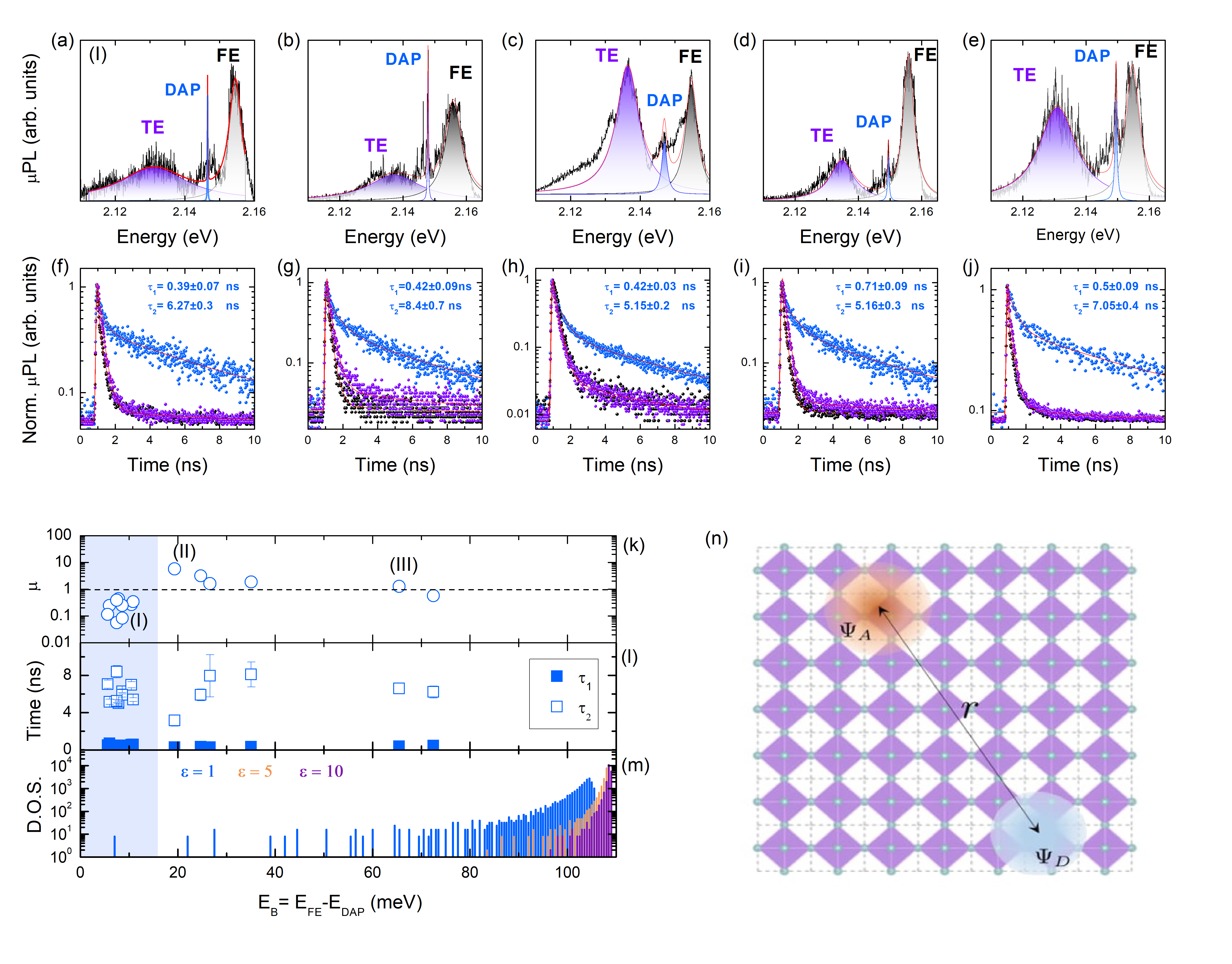}
	\caption{Donor Acceptor Pair (DAP) optical emission signatures in n = 2 phase samples. (a-e) $\mu$-PL spectra of five different samples with highlighted free exciton emission peak (FE, black color), self-trapped emission peak (TE, violet color), and ionized Donor-Acceptor pair emission peaks (DAP, blue color). (f-j) Corresponding $\mu$-TRPL traces of the observed FE (black scattered dots), TE (violet scattered dots), and DAP (blue scattered dots) optical transitions, which are displayed from left-hand to right-hand as the DAP binding energy is reduced. (k) Statistics of the relative intensity between both fast and slow radiative channels, ($\mu$) as a function of its binding energy for the DAP transition. ((I) is linked to \ref{Fig_3}.a, and the $\mu$-PL peaks associated with (II) and (III), are shown in Figure S1 in Supp. Info.) (l) Statistics of the fast ($tau_{1}$) and slow ($\tau_{2}$) decay times as a function of the binding energy of the DAP state. (m) Model output of the DAP density of states (DOS) as a function of binding energy, in logarithmic scale for three different dielectric constants. (n) Sketch of the DAP binding complex states in a 2D perovskite lattice.}
	\label{Fig_3}
\end{figure}

    \subsection*{Donor-Acceptor pair optical emission}
    
    The higher complexity that is observed in both FWHM and decay time statistics in this n=2 phase is accompanied in some cases with the presence of a third and narrow optical transition, in which central peak energy position ranges between TE and FE optical transitions (see Figure \ref{Fig_3}.a-e). Here we present the spectral and decay time characteristics of this third optical transition. We label it as the Donor-Acceptor Pair (DAP) state. Later on, we will provide solid experimental evidence to support this assignation. Figures \ref{Fig_3}.a-e show $\mu$-PL spectra of n = 2 phase samples, where the DAP transition is present. We performed multi-Lorentzian fitting in each spectrum. The contribution of FE and TE optical emission is shown with black and violet shadows. Blue shadows indicate the emission from DAP transition, with characteristic FWHMs ranging from 150 $\mu$eV up to $\simeq$ 5 meV. The spectra are arranged from left-hand to right-hand  to show DAP transition with decreasing binding energy, which is here defined as $E_{B} = E_{FE} - E_{DAP}$. FWHM \& $E_{B}$ pairs take the following values---a: 0.12 \& 8.1 meV, b: 0.3 \& 8.0 meV, c: 1.8 \& 7.75 meV d: 0.55 \& 6.4 meV and e: 0.8 \& 5.26 meV (as shown by the statistics of DAP peak linewidth FWHM as a function of $E_{B}$ in Figure S2 in Supp. Info.). As before, we performed $\mu$-TRPL measurements of FE, TE and DAP transitions. Figures \ref{Fig_3}.f-j show the FE and TE time traces following the same evolution described earlier; that is, fast decay dominating over a second slow decay (here, FE and TE are plotted with black and violet scattered dots). However, the sharp DAP transition present in this phase shows quite different behaviour. The DAP dynamics (plotted with blue scattered dots) also contain both fast and slow decay times. Nonetheless, here the slow decay represents the dominant intensity contribution. Figure \ref{Fig_3}.k shows the evolution of $\mu$ as a function of $E_{B}$. As can be seen, there is a clear energy range where the radiated intensity from the slow decay channel  clearly dominates over the fast decay ($\mu < 1$), where this last fast decay contribution is almost negligible; this energy range is marked with a bluish shadow in figure \ref{Fig_3}.k-m (dot label with (I) is linked to the spectra shown in figure \ref{Fig_3}.a, and the $\mu$-PL peaks associated with (II) and (III), are shown in Figure S1 in Supp. Info.). The residual presence of the fast decay when filtering the DAP transition can be associated with the background emission from FE and TE transitions. Figure \ref{Fig_3}.a-e shows how the Lorentzian tail of FE and TE produces an overlap with the DAP peak energy position, and hence it is plausible that $\mu$-TRPL traces contain some contribution from the FE and TE tails. Figure \ref{Fig_3}.l shows a statistical sample of DAP $\tau_{1}$ and $\tau_{2}$ values as a function of $E_{B}$. We can assume that DAP optical transition is characterized by longer time dynamics, which is in correspondence with its sharpest spectral features. However, neither decay and FWHM are Fourier transform related, and hence we should expect the effect of extra carrier dynamics mechanism in the DAP recombination, even though our limited spectral resolution ($\simeq 100 \mu eV$) reduces our capacity to estimate the real FWHM values. To obtain the full description of the characteristics and magnitude of the correlation time and dephasing process acting on the DAP transition, it will be necessary to perform alternative interferometric measurements to derive its first order coherence function \cite{berthelot2006unconventional}. 

    \section*{Discussion}
    
    With the intention of evaluating our starting hypothesis of addressing these $\mu$-PL and $\mu$-TRPL special features found in the n = 2 phase with the DAP state, we first compare our optical emission results with a simple model to calculate the DAP Density of States as a function of its binding energy in 2D perovskites. Following this procedure, we will estimate the possibility to find sharp-peaked transitions along the energy range between FE and TE. Second, we carried out dedicated high resolution photoemission experiments and $\mu$-Raman measurements to check the validity of these assumptions and to identify the electronic states that originate the DAP transition. 
    
    To develop our model, we assume the existence of donor (D) and acceptor (A) states within the bandgap. According to Ref. \cite{Thomas1964}, the substitution defects of MA and Pb leads to D and A states below and above the conduction and valence band edges of approximately 0.15 eV.\cite{Wu2015} The energy of the emitted photon in a DAP transition is given by:
    $$E_{SP}(r) = E_{FE} - E_{A-} - E_{D+} + \frac{e^2}{\varepsilon r}$$
    where $r$ is the distance between donors and acceptors. The term $\frac{e^2}{\varepsilon r}$ accounts for the energy correction due to the Coulomb interaction between the ionized D$^+$A$^{-}$ pair and the corresponding increase of the photon energy. Figure \ref{Fig_3}.n shows the geometry of the square lattice of the 2D perovskites with a sketch of the D$^+$A$^{-}$ bound state. For the calculations, we have assumed a two-layers 2D perovskite, trying to emulate the case with phase n = 2. The density of states (DOS) is obtained by calculating $E_{SP}$ for all possible distance between an electron in the donor site and the hole in the acceptor site. Figure \ref{Fig_3}.m shows the DOS for three different values of the dielectric screening $\varepsilon$. The first situation, $\varepsilon=1$ represents the unscreened situation, which is typical of molecules. We appreciate large exciton binding energies with a discrete energy spectra. In the case of a larger dielectric screening ($\varepsilon=5,10$), the exciton DOS tends to collapse close to the TE peak and the discrete energy spectra is not that evident. In the realistic case of 2D perovskites made of one and two layers, dielectric screening is weak and strongly anisotropic \cite{Molina-Sanchez2018} in the out-of-plane direction, exhibiting large exciton-binding energies. As evidenced by the DAP peaks with energy position close to the FE peak, the dielectric screening of the studied samples would be between 1 and 5, showing discrete DAP peaks for the range between FE and TE transitions. We can assume slow decays times of the DAP complexes because its wavefunctions counterparts will not overlap very efficiently, due to their large spatial separation. Our model predicts the existence of isolated DAP states in the vicinity of the FE state, which supports the assumption of narrow DAP transitions. The decay time of DAP transitions is fundamentally determined by the donor-acceptor separation. This is analogous to the spatially separated interlayer excitons in the heterobilayers of 2D materials.\cite{Torun2018} Ab initio calculations and spectroscopy measurements report that the decay times for interlayer excitons are in the order of nanoseconds \cite{Jiang2021}, which is   in agreement with the DAP transitions of our work.

\begin{figure}[h]
	\centering
	\includegraphics[scale=0.42]{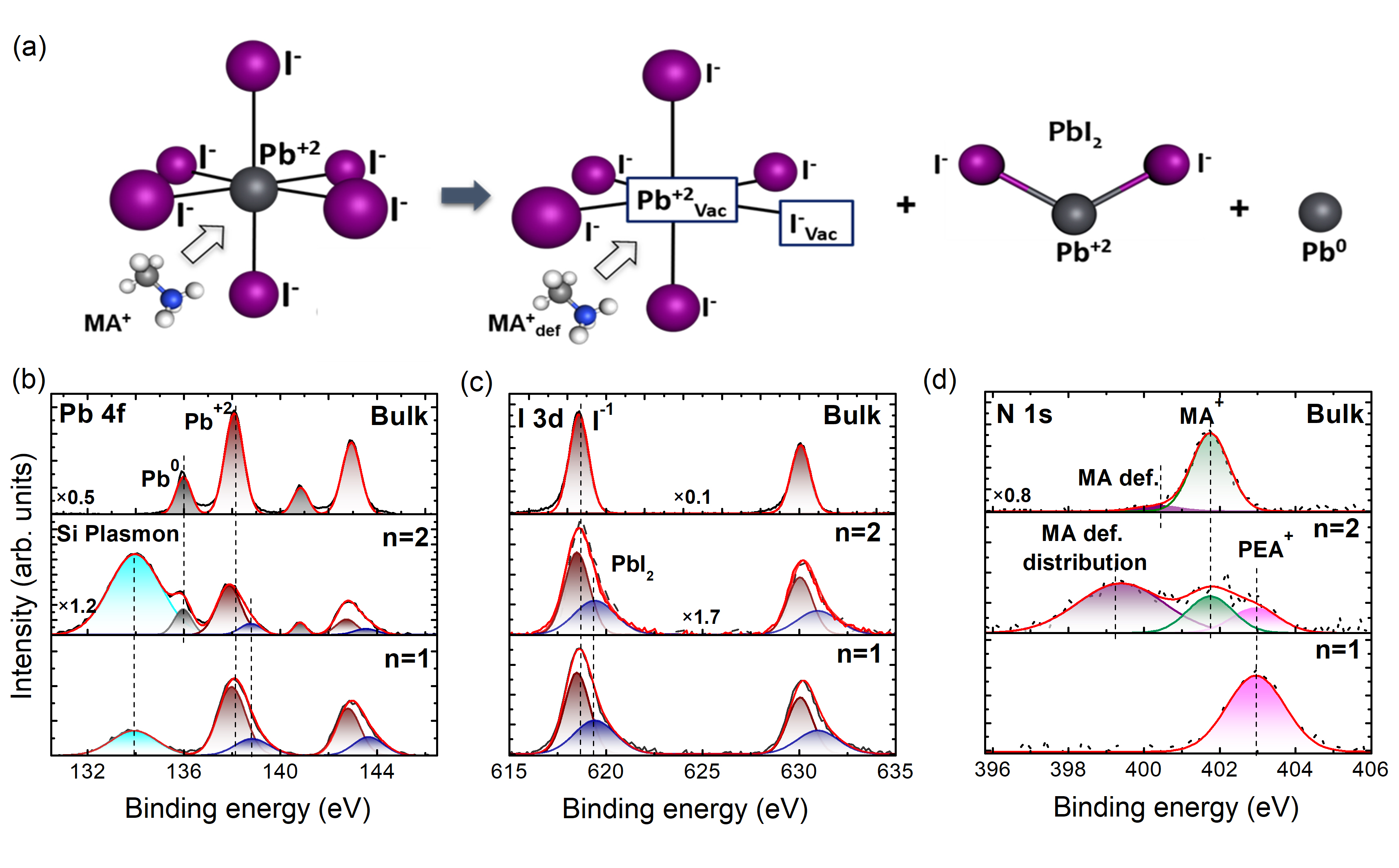}
	\caption{(a) Scheme of the non-defective and defective $PbI_{6}$ octahedra for n = 2 HPs with highlighted Pb and I vacancies, which might lead to the formation of $PbI_{2}$ molecule or metallic $Pb^{0}$. Created Pb vacancy might be then occupied by methylammonium (MA) molecule. High energy resolution XPS of (b) Pb 4f , (c) I 3d  and (d) N 1s core-levels measured in n = 1, n = 2 and n = ∞. 
	  The broad peak appearing on the low binding energy side of Pb 4f spectra for n = 1 and 2 samples (shaded in light blue), corresponds to the second replica of the Si 2p plasmon (plasmon energy of $\sim$ 17.8$~eV$ from Si 2p$_{3/2}$).}  
	\label{Fig_4}
\end{figure}
    
    With the aim to disentangle the origin of electronic states responsible for DAP transitions, we have performed high-resolution X-ray photoelectron spectroscopy (HR-XPS) and $\mu$-Raman spectroscopy measurements in these perovskites. Figures \ref{Fig_4}.b-d show the Pb 4f, I 3d, and N 1s core level spectra acquired in both phases n = 1, n = 2 and in a reference bulk sample. These measurements are expected to provide for direct information about the oxidation degree and chemical environment of each atomic species. In the bulk samples, Pb 4f and I 3d core-level spectra are dominated by $Pb^{+2}$ and $I^{-1}$ spin-orbit doublets with a $Pb 4f_{7/2}$ component located at 138.1 eV and a $I 3d_{5/2}$ component at 618.7 eV, respectively. These components are attributable to Pb and I atoms in the inorganic octahedral anions. 
    Together with the main $Pb^{+2}$ core-level signal, we have observed traces from $Pb^{0}$ (with its $Pb 4f_{7/2}$ component located at 135.9 eV) that reveals the presence of metallic Pb segregated to the sample surface. The Pb 4f and I 3d core-level spectra acquired in the n=2 sample exhibit also traces from $Pb^{+2}$ and $Pb^{0}$ species. 
  However, we have detected additional Pb 4f and I 3d doublets, these appearing at the high energy side of the main $Pb^{+2}$ and $I^{-1}$ doublets with their $Pb 4f_{7/2}$ and $I 3d_{5/2}$ components located at 138.9 eV and 619.5 eV, respectively. This result, together with the fact that these new Pb 4f and I 3d components exhibit an atomic Pb/I ratio of $\sim$0.3$\pm$0.1, suggests that their origin may be related to the presence of $PbI_{2}$-like species at the sample surface ($\mu$-Raman spectroscopy results shown in Figure S3 of Supplementary Information also support this assignment). In the Pb 4f and I 3d core-levels measured in the last kind of samples, the n=1 ones, spectra reveal the presence of the $PbI_{2}$-like species accompanying to the Pb and I atoms in the inorganic octahedral anions. However, no traces from metallic $Pb^{0}$ have been observed, interestingly.

    These photoemission results reveal a different chemical behavior as going from one kind of samples to the other, which imply a rich structural-dependent chemical activity. In all kind of samples, Pb 4f and I 3d spectra appear dominated by signals coming from the inorganic octahedral sheets. However, we have detected metallic $Pb^{0}$, but not $PbI_{2}$-like species, in the bulk samples and we have detected $PbI_{2}$-like species, but not metallic $Pb^{0}$, in the n=1 samples. Only in the n=2 samples both metallic $Pb^{0}$ and $PbI_{2}$-like species coexist. Structurally, these samples differ one from each other, precisely, in the nature of their organic components: $MA^{+}$ cations constitute the organic species in bulk samples, whereas $PEA_{2}$ ones are those of the n=1 samples. These facts suggest that different organic components give rise to defects of different chemical nature: $MA^{+}$ cations somehow promote segregation of metallic $Pb^{0}$ whereas $PEA_{2}$ cations tend to favor decomposition of inorganic sheets to produce $PbI_{2}$-like species.

    Perovskites can naturally host a wide variety of optically active defects whose optical recombination energy depends on their chemical nature \cite{lekina2019excitonic}. For instance, several defects acting as acceptors ($I_{i}$, $MA_{Pb}$, $V_{MA}$, $V_{Pb}$, $I_{MA}$, $I_{Pb}$) or donors ($MA_{i}$, $Pb_{MA}$, $V_{I}$, $Pb_{i}$, $MA_{I}$, $Pb_{I}$) have been found in 3D \cite{wang2018defects, merdasa2017supertrap, yin2014unusual} and 2D \cite{xiao2016defect} LHPs (here, V = vacancy, subscript i =  interstitial, and $A_{B}$ denotes B substituted by A) \cite{merdasa2017supertrap}. Taking this into account, photoemission results reported here evidence a n-dependent chemical activity in 2D perovskites, which would give rise to optically active defects whose nature would depend on the specific organic component involved in. This fact becomes particularly relevant for disetangling optical properties of 2D perovskites with n=2, in which both $PEA_{2}$ and $MA^{+}$ organic cations coexist in a similar proportion. In fact, we believe that a joined effect of these two organic cations seems to be responsible for the generation of DAP in 2D perovskites with n=2: $MA^{+}$ cations seem to promote Pb segregation, but the presence of $PEA_{2}$ cations, which strongly disturb inorganic sheet, collaterally enhances Pb segregation favouring that $MA^{+}$ cations to occupy Pb atomic sites within the octahedrals and creating $MA_{Pb}$ acceptors.\cite{xiao2016defect} At the same time, displaced Pb atoms may occupy $MA^{+}$ sites out of the octahedrals, eventually acting as $Pb_{MA}$ donors. It seems reasonable to assume that the joint action of both cations affects to the same octahedral anion, then, it would be expected that $MA_{Pb}$-related acceptors and $Pb_{MA}$-related donors are close to each other. Consequently, their wavefunctions would easily interact giving rise to DAPs.

    The N 1s core-level spectra Figures \ref{Fig_4}.d, as measured by HR-XPS in 2D perovskites, seem to support the scenario proposed above. In our samples, the N 1s signal stems from the organic components of the 2D perovskites. In  fact, in the bulk and in the 2D perovskite with n=1, the dominant N 1s singlet signal comes from the $MA^{+}$ and $PEA_{2}$ cations, respectively. Only an additional weak signal can be observed at the low energy side of the N 1s spectrum measured in the bulk sample, which may be attributed to $MA^{+}$ cations in Pb sites (hereafter called the $MA^{+}$ defects).
  The N 1s spectrum acquired in the 2D perovskite with n=2 exhibits both peaks coming from the $MA^{+}$ and $PEA_{2}$ cations located at their own crystal sites.
     In addition, a new component appears, which is correlated with that of PbI$_{2}$~ in the Pb 4f spectrum and corresponds to the aforementioned $MA^{+}$ defects. In the n = 2 sample, the N 1s contribution of $MA^{+}$ defects is dominant, indicating that PEA$_{2}$~ promotes the exchange of $MA^{+}$~ cations with Pb in octahedral sites.

    In conclusion, our work shows that the research on DAP in 2D materials represents a growing field with high potential. Our observation of the narrow and discrete PL lines associated to DAP transitions is added to recent reported properties, such as exciton binding energy tunability by donor acceptor interactions\cite{passarelli2020tunable} and DAP single photon emission performance in 2D hBN\cite{tan2021donor}. 
       By means of HR-XPS spectroscopy, we show that the two organic components present in the n=2 sample promote the formation of the defects responsible of the DAP transition.
       In detail, the coexistence of $MA^{+}$ cation, giving rise to Pb segregation, and $PEA_{2}$ cation, causing the inorganic lattice sheet distortion, might favour the displacement and exchange of $MA^{+}$ and Pb atoms, hence the formation of $MA_{Pb}$ acceptor and $Pb_{MA}$ donor states.
       Our model predicts the presence of low density DAP states in the spectral range close to the FE emission, and hence both low DOS and large donor acceptor wavefunction separation explain the measured DAP narrow optical transitions with nanosecond lifetimes. To provide a future application based on the control of these states,  specific sample preparation routes should be designed to control and optimize the DAP state population, localization, or isolation. Due to its featured spectral properties, along with their high photoluminescence brightness, the DAP state in 2D Perovskites represents an interesting electronic state to be considered for the development of the emerging field of quantum technologies, such as quantum polaritonics \cite{ballarini2019polaritonics} or quantum non-linear optics \cite{chang2014quantum}.

\section*{Materials and Methods}

\subsection*{Synthesis:}
Materials were prepared by modifying a synthetic procedure that has been reported elsewhere \cite{stoumpos2016ruddlesden}. Synthesis of  PEA$_{2}$PbI$_{4}$ (n=1).  Solution A. PbO powder (1116 mg, 5 mmol) was dissolved in a mixture of 57\%
w/w aqueous HI solution (5.0 mL, 38 mmol) and 50$\%$ aqueous H$_3$PO$_2$ (0.85 mL, 7.75 mmol) by heating to boiling under vigorous stirring for about 10 min. It was then possible to see the formation of a bright yellow solution. Solution B. In a separate beaker, C$_{6}$H$_{5}$CH$_{2}$CH$_{2}$NH$_{2}$ (phenetylamine, PEA) (628  $\mu$L, 5 mmol) was neutralized with HI 57\% w/w (2.5 mL, 19 mmol) in an ice bath, resulting in a clear pale-yellow solution. If a solid precipitate or the formation of a suspension is observed, then it is possible to heat slightly until it has dissolved. Once both solutions have been prepared, we proceed to the slow addition of Solution B to Solution A. A quick addition could produce a precipitate, which was subsequently dissolved as the combined solution was heated to boiling. After keep the solution at  boiling point for 10 min, the stirring was then discontinued and the solution was left to cool to room temperature, during which time orange rectangular-shaped plates started to crystallize. The crystallization was deemed to be complete after 1 hour. The crystals were isolated by suction filtration and thoroughly dried under reduced pressure.  The obtained single crystals were washed with cold diethyl ether.

Synthesis of PEA$_{2}$MAPb$_{2}$I$_{7}$ (n=2).  Solution A. PbO powder (1116 mg, 5 mmol) was dissolved in a mixture of 57\% w/w aqueous HI solution (5.0 mL, 38 mmol)  and 50\% aqueous H$_{3}$PO$_{2}$ (0.85 mL, 7.75 mmol) by heating to boiling under vigorous stirring for about 5 min and forming a yellow solution. Then, CH$_{3}$NH$_{3}$Cl powder (169 mg, 2.5 mmol) was added very slowly to the hot yellow solution, causing a black precipitate to form, which rapidly redissolved under stirring to achieve a clear bright yellow solution. Solution B. In a separate beaker, C$_6$H$_5$CH$_2$CH$_2$NH$_2$ (PEA) (880  $\mu$L, 7 mmol) was neutralized with HI 57 \% w/w w/w (2.5 mL, 19 mmol)  in an ice bath, resulting in a clear pale-yellow solution. The addition of the PEAI solution to the PbI$_{2}$ solution initially produce a black precipitate, which was dissolved under heating the combined solution to boiling. After discontinuing the stirring, the solution was cooled to room temperature, and cherry red crystals started to emerge. The precipitation was deemed to be complete after $\sim$1 h. The crystals were isolated by suction filtration and thoroughly dried under reduced pressure. The material was washed with cold diethyl ether.

\subsection*{Crystal cleaving and sample preparation:}
A large statistical set of 2D PVKs and 3D single crystals was investigated. All of the measurements are carried out at room temperature. 2D PVKs crystals (PEA$_{2}$PbI$_{4}$ and PEA$_{2}$MAPb$_{4}$I$_{7}$) have been mechanically exfoliated on Si substrates by scotch tape.

\subsection*{Optical spectroscopy \& Confocal optical microscopy:}
All low-temperature measurements were carried out using a standard $\mu$-PL setup, with the samples placed in the cold finger of a vibration-free closed-cycle cryostat (AttoDRY800 from Attocube AG).  A continuous-wave (pulsed) excitation laser with a wavelength of $\lambda$ = 405 nm (450 nm) was used to excite the sample, providing spectral $\mu$-TRPL measurements. Furthermore, single-mode fiber-coupled laser diodes at $\lambda$ = 532 nm and 660 nm were used to align the collection and excitation spots for 2D perovskite (n=1, n=2), respectively. The confocal pinholes will be defined by the single-mode optical fibers used for excitation and detection. Excitation and detection were carried out using a 50X microscope objective with a long working distance and a numerical aperture of NA = 0.5, which was placed outside the cryostat. The sample's emission was long-pass filtered, dispersed by a double 0.3 m focal length grating spectrograph (Acton SP-300i from Princeton Instruments), and detected with a cooled Si CCD camera (Newton EMCCD from ANDOR) for recording $\mu$-PL spectral and a silicon single-photon avalanche photodiode detector (from Micro Photon Devices) connected to a time-correlated single-photon counting electronic board (TCC900 from Edinburgh Instruments) for $\mu$-TRPL measurements \cite{adl2021homogeneous}.

\subsection*{Raman and photoemission:}

Micro-Raman spectroscopy using a confocal Raman microscope Horiba-MTB Xplora was carried out using a red laser illumination with 638 nm excitation wavelength, 0.8 – 8 mW laser power, and 100X 
objective (N.A.= 0.9, WD = 0.21 mm). 
Part of the high-resolution x-ray photoemission (HR-XPS) experiments were performed at the ALOISA beamline of the Elettra Sincrotrone of Trieste, Italy.\cite{Floreano2008ALOISA}  In this case, the X-ray photon energy of 515 eV (overall resolution of 150 meV, FWHM) was used for C 1s, N 1s and Pb 4f core-levels and photon energy of 740 eV (overall resolution of 280 meV) was used for I 3d core-levels.
The samples (n = 1 and 2) were kept at a grazing angle of $4^{\circ}$~ in close to p-polarization.
The spectra were measured in normal emission with a homemade hemispherical spectrometer (mean radius of 66 mm) equipped with the Elettra 2D delay-line detector.   
The other part of the HR-XPS experiments was performed in a SPECS GmbH system (base pressure 1.0$\times$10$^{-10}$ mbar) equipped with a PHOIBOS 150 2D-CMOS hemispherical analyzer. Photoelectrons were excited with the Al-K$_{\alpha}$ line (1486.7 eV) of a monochromatic x-ray source $\mu$-FOCUS 500 (SPECS GmbH). Measurements were taken at room temperature with a pass-energy of 20 eV. For all of the HR-XPS measurements, freshly exfoliated 2D PVKs were exposed to air and placed in ultra-high vacuum (UHV) conditions (with an analysis chamber base pressure lower than 1.0$\times$10$^{-10}$ mbar. In contrast, 3D samples were just grown and immediately introduced into the UHV chamber to avoid any perturbation of the surface species induced from being in contact with standard atmospheric conditions.

\section*{Acknowledgments}

This work was made possible by the 2D-SPD(Two-Dimensional Semiconductor Photonic Dots) project, funded by the Spanish Ministry of Science MICINN, AEI (RTI2018-099015-J-I00) and the European Regional Development Fund (ERDF), the Horizon 2020 research and innovation program of the EU through the S2QUIP project (grant agreement No. 8204023), the Ministerio de Ciencia e Innovación, which is part of Agencia Estatal de Investigación (AEI), through the project PID2020-112507GB-I00 (Novel quantum states in heterostructures of 2D materials) and PID2019-107314RB-I00 (Stable), and the Generalitat Valenciana through the project PROMETEU/2021/082. 
The research leading to this result has been supported by the project CALIPSOplus under Grant Agreement 730872 from the EU Framework Programme for Research and Innovation HORIZON 2020.
A. M.-S. acknowledges the Ram\'on y Cajal programme (grant RYC2018-024024-I; MINECO, Spain). M. K. acknowledges the APOSTD programme (contract APOSTD/2020/103).

\bibliographystyle{unsrt}
\bibliography{Bibliography_edited}

\end{document}